\documentclass[11pt]{article}

\usepackage{color}

\usepackage{amssymb,stmaryrd,amsthm}

\usepackage{amsmath,amsbsy}

\usepackage{MnSymbol,natbib}

\usepackage{graphics,graphicx,hyperref}
\usepackage[export]{adjustbox}

\newcommand{\bdm}{\begin{displaymath}}
\newcommand{\edm}{\end{displaymath}}

\newcommand{\beq}{\begin{equation}}
\newcommand{\eeq}{\end{equation}}

\newcommand{\bea}{\begin{eqnarray}}
\newcommand{\eea}{\end{eqnarray}}

\newcommand{\beas}{\begin{eqnarray*}}
\newcommand{\eeas}{\end{eqnarray*}}

\newcommand{\ttt}[1]{\texttt{#1}}

\newcommand{\red}{\begin{color}{red}}

\newcommand{\redd}{\end{color}}

\newcommand{\green}{\begin{color}{green}}

\newcommand{\greenn}{\end{color}}

\newcommand{\bcx}{\begin{color}{black}}

\newcommand{\ec}{\end{color}}

\title{Reproducibility and Statistical Methodology}
\author{Anthony Almudevar, PhD  \\ Department of Biostatistics and Computational Biology, \\ University of Rochester,  Rochester NY\\ \\
Jacob Almudevar, MSc \\ Department of Mathematics and Statistics, \\ University of New Hampshire, Durham NH}

\begin{document}

\maketitle

\noindent\textbf{Abstract.}
In 2015 the Open Science Collaboration (OSC) (Nosek et al 2015) published a highly influential paper which claimed that a large fraction of published results in the psychological sciences were not reproducible. In this article we review this claim from several points of view. We first offer an extended analysis of the methods used in that study. We show that the OSC methodology induces  a bias that is able by itself to explain the discrepancy between the OSC estimates of reproducibility  and other more optimistic estimates made by similar studies. 

The article also offers a more general literature review and discussion of  reproducibility  in experimental science.  We argue, for both scientific and ethical reasons, that a considered balance of false positive and false negative rates is preferable to a single-minded concentration on false positive rates alone. 


\section{Introduction}

The value of any kind of research depends on the reproducibility of the results. ``The salutary habit,'' suggests Ronald  Fisher, ``of repeating important experiments, or of carrying out original observations in replicate, shows a tacit appreciation of the fact that the object of our study is not the individual result, but the population of possibilities of which we do our best to make our experiments representative.'' \citep{fisher1925}. 

If a study produces an outcome which cannot be reproduced in a similar environment at another point in time, then such a study has failed in its original goal to provide observations that allude to real-life phenomena. Therefore, ensuring the reproducibility of all studies is a top priority in all fields of science. However, \bcx a number of investigations \ec into the current climate of scientific studies have suggested that a large number of published results are unable to be reproduced, implying that these results have no implications outside of the initial test conditions. This issue is known in the literature as the \bcx \textit{reproducibility} (or \textit{replication}) \textit{crisis}, \ec and has become a significant concern.  \bcx In fact, according to a survey of researchers reported in \cite{baker2016}  `` ... 52\% of those surveyed agree that there is a significant ‘crisis’ of reproducibility ...''. \ec  \bcx As a consequence,  \ec failure of reproducibility is seen as a major issue across most fields of science, with many individuals and organizations investigating these claims. 

In order to test this, the Open Science Collaboration (OSC) (\texttt{osf.io/vmrgu}) \bcx conducted a replication project (OSC-RP),  selecting \ec 100 studies from prominent psychological journals and \bcx reproducing \ec them in close to original conditions \citep{open2015estimating}. The OSC-RP reports that while $97\%$ of the original experiments were shown to have significant results \bcx ($P \leq 0.05$), \ec of these only 36\% of the reproduced experiments were found to be significant under \bcx similar \ec conditions. Furthermore, $47\%$ of effect sizes from the original experiments were within the $95\%$ confidence interval of the effect sizes from reproduced studies, and $39\%$ of effects were subjectively considered to be successful reproductions.

The OSC-RP report does not discern any particular systematic flaws in experimental or statistical methodology, although it conjectures issues regarding incentive in the scientific community. The OSC-RP does conclude, however, and implies as such in their report, that the scientific community faces a self-evident risk of irreproducible and unreliable experimental data from major scientific journals. If true, this is a serious concern that requires some degree of reform in experimentation and publication practices.

However, not all studies of reproducibility yield pessimistic conclusions. \cite{etz2016bayesian} presents the notion that the reported failure of reproducibility from the OSC-RP's report can be attributed to an ``overestimation of effect sizes''. \bcx \cite{klein2014}   summarizes \ec a separate replication study, the  Many Labs Replication Project (ML-RP), which reported a much higher reproducibility rate \bcx of $85\%$,  \ec one compatible with strict adherence to good experimental and statistical practice  (\url{https://osf.io/wx7ck/k/}). \bcx In a similar study reported in \cite{camerer2016} 18 experimental studies  in the field of economics were replicated, again yielding a higher reproducibility rate than the OSC-RP (61\%), while using a similar replication protocol (ECO-RP).  See also the exchange in \cite{gilbert2016} and \cite{anderson2016} following publication of \cite{open2015estimating}.  \ec
 
Therefore, there may be a different conclusion to discern from the statistical data presented by the OSC-RP, one that is both sensible and elementary in regards to basic statistical principles. In order to investigate further, we must begin by clarifying and reconsidering the degree of reproducibility that researchers should expect from such studies.

\section{A simple mathematical model for reproducibility}\label{sec.model}

We first develop a \bcx ``reproducibility model''  \ec with which to precisely define a reproducibility rate and to offer guidance regarding  its estimation. Throughout, model assumptions will never deviate from standard practice, and we will assume, in particular, that probabilities of conventional type I error (false positive) and type II error (false negative) are  accurately reported where needed. If this model is able to predict the reproducibility rates reported by the OSC-RP under these conditions, then it provides no basis to fault  contemporary research and publication practice.    


\subsection{Model definition}\label{sec.model.def}

\begin{figure}[h] 
\centering
\includegraphics[width=4.5in,height=2.5in,viewport=20 120 830 600, clip, frame]{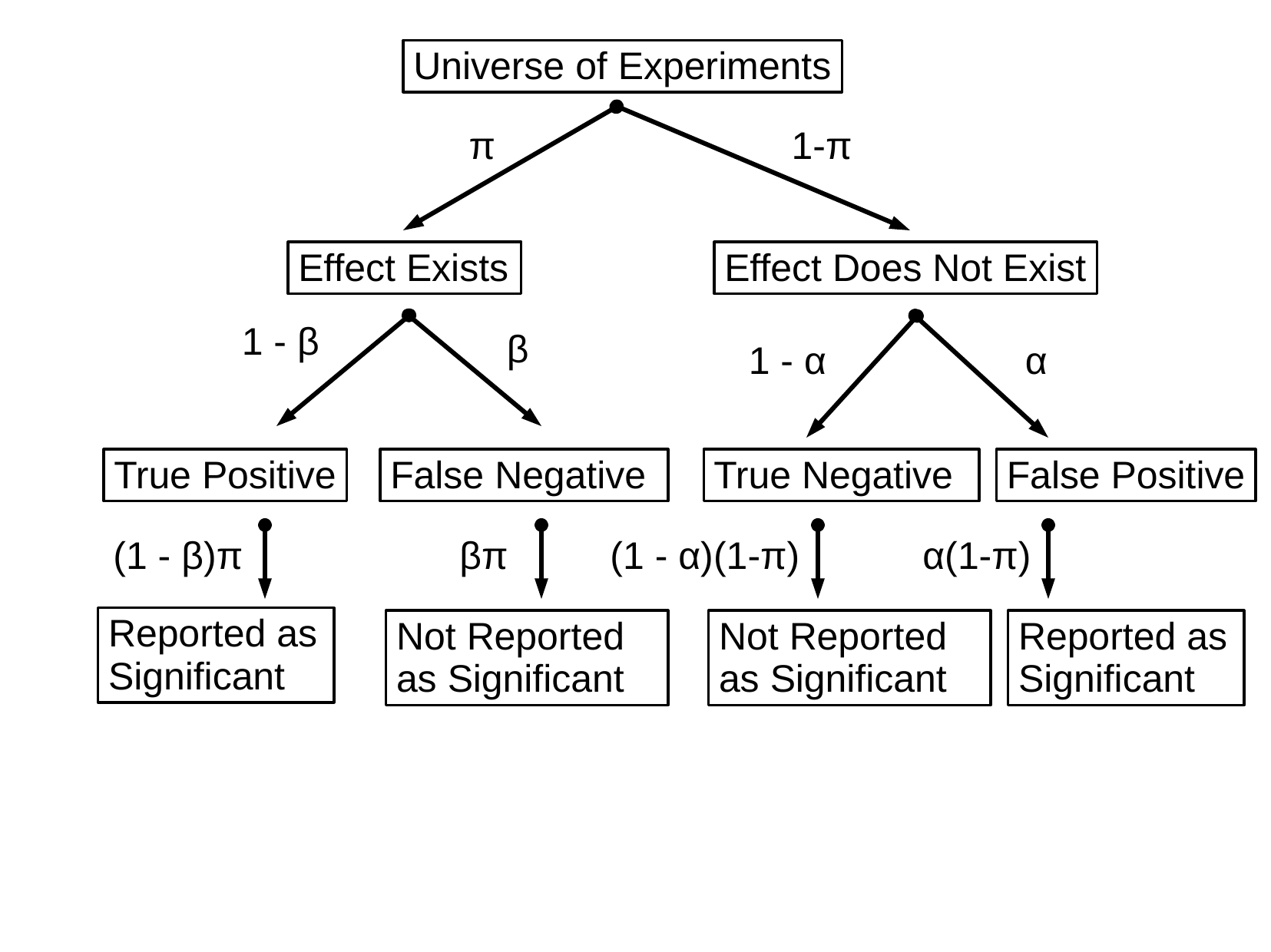}
\caption{Decision tree representation of reproducibility model.}\label{fig.decision.tree}  
\end{figure}

Let us define a universe $\mathcal{U}$ of hypothesis tests, for which we have null hypothesis $H_o$ and alternative hypothesis $H_a$. \bcx Alternative hypotheses represent effects of  scientific interest, and as a consequence are published in journals. A proportion $\pi$ of $\mathcal{U}$, which we refer to as \emph{effect prevalence}, are truly $H_a$. \ec Let $\alpha$ represent the type I error and let $\beta$ represent the type II error. All possible reported outcomes of a study can be represented with \bcx the decision tree illustrated in Figure \ref{fig.decision.tree}. 
\ec

Let $A$ represent the event that $H_a$ is true, and let $E$ represent the event that the study produces a positive test result, in the form of the $P$-value threshold $P \leq \alpha$. The \emph{positive predictive value} ($PPV$) can be taken as $PPV=P(A \mid E)$. Using Bayes' theorem we obtain, in terms of the odds,
\begin{equation} \label{Odds(PPV)}
\textup{Odds}(PPV)=\bigg(\frac{P(E \mid A)}{1-P(E^c \mid A^c)}\bigg)\cdot\textup{Odds}\big(P(A)\big).
\end{equation}
Relative to these definitions, we may define $\alpha=P(E \mid A^c)$ and $\beta=P(E^c \mid A)$. Inserting these values into Equation (\ref{Odds(PPV)}), we obtain:
\begin{equation} \label{Odds(PPV)2}
\textup{Odds}(PPV)=\bigg(\frac{1-\beta}{\alpha}\bigg)\cdot\textup{Odds}(\pi).
\end{equation}
Assuming typical  values $\alpha=0.05$ and $\beta=0.1$, this would suggest that the odds that a positive is a true positive are about $18$ times the odds that $H_a$ is true. 

$PPV$ is a fair measure of reproducibility; any value less than 1 implies that test results may not be accurate. Unfortunately, directly calculating the true value of $PPV$ presents \bcx practical difficulties. \ec Thus, we present a definition for a value $PPV_{obs}$ that represents the \textit{observed} reproducibility in a replication study \bcx (assuming all studies used report significant effects). \ec Suppose such a study reports type I error $\alpha^*$ and type II error $\beta^*$, which need not equal those of the original study. Then $PPV_{obs}$ has the following \bcx relationship to $PPV$:\ec
\begin{equation} \label{PPVobs}
PPV_{obs}\approx PPV(1-\beta^*)+(1-PPV)\alpha^*.
\end{equation}
\bcx It is important to note that \ec the distinction between $PPV$ and $PPV_{obs}$ is dependent only on the protocol of the \bcx replication study,  so that we can \ec control for $\alpha^*$ and $\beta^*$ in order to estimate $PPV$. From Equation \eqref{PPVobs} we have 
\begin{equation} \label{PPV*}
PPV\approx\frac{PPV_{obs}-\alpha^*}{1-\alpha^*-\beta^*}.
\end{equation}

Thus, together, Equations  \eqref{Odds(PPV)2}-\eqref{PPVobs} can be used both to define what an ideal reproducibility rate should be, and to quantify any deviation from that ideal. Assuming the values  $\alpha^*$ and $\beta^*$ associated with a reproducibility protocol are accurately reported, any such discrepancy can be quantitatively related to differences in nominal and actual values of $\alpha$ or $\beta$, or to an overly optimistic estimate of the effect prevalence $\pi$ itself. 


\subsection{What value can we expect effect prevalence $\pi$ to be?}\label{sec.prevalence}

It is the value of $\pi$ that interests us the most, as it leads to an important observation from the OSC-RP's original analysis of reproducibility (emphasis added):
\begin{quote}
On the basis of only the average replication power of the $97$ original, significant effects [$M=0.92$, median ($Mdn$) $=0.95$], we would expect approximately $89$ positive results in the replication $\textbf{if all original effects were true and accurately estimated}$; however, there were just $35$ [$36.1\%$; $95\%$ $\textup{CI}=(26.6\%,46.2\%)$], a significant reduction [McNemar test, $\chi^2(1)=59.1,P<0.001$]. \citep{open2015estimating}. 
\end{quote}
Given the relations defined in  Equations \eqref{Odds(PPV)2} and \eqref{PPVobs}, this suggests that the authors are assuming $PPV=1$ and, therefore, $\pi=1$. \bcx This cannot be the case; if it were, we could simply assume $H_a$ was true in all cases, and no experiment would have to be performed. \ec

\bcx
Otherwise, what should we expect $\pi$ to be?   Identifying a realistic value involves clarifying our definition of  the population  of hypothesis tests $\mathcal{U}$. \textit{Primary analyses} involve the resolution of  a hypothesis of scientific consequence, in which type II error $\beta$ is controlled to be a commonly used value such 0.1.   \textit{Secondary analyses} report effects  that enhance a primary analysis. For this type of analysis, controlling $\beta$ is not considered as essential or practical.  Hypotheses presented in primary analyses are typically supported by statistical and mechanistic prior evidence. We therefore conjecture that $\pi$ should be smaller for secondary analyses, and will more generally depend on particular features of the type of research conducted. Similarly, we would expect $\pi$ to be small for \textit{exploratory studies}, such as those involving a search for a significant treatment effect within high-throughput forms of data. 

We can use our reproducibility model to predict $PPV$ or $PPV_{obs}$  for various scenarios. Table \ref{table.ppv.pred} lists some examples, assuming a replication protocol with errors $\alpha^* = \alpha = 0.05$ and $\beta^* = \beta = 0.1$. 
\ec

\begin{table}[ht]
\centering
\caption{Predicted values of $PPV$ and $PPV_{obs}$ based on Equations  \eqref{Odds(PPV)2}-\eqref{PPVobs} for varying prevalence values of $\pi$, assuming a replication protocol with type I and II errors $\alpha^* = \alpha = 0.05$ and $\beta^* = \beta = 0.1$.}\label{table.ppv.pred} 
\begin{tabular}{|c|l|c|c|r|}
\hline
Case &Research environment & $\pi$ & $PPV$ & $PPV_{obs}$ \\ \hline
1&Overly optimistic (OSC-RP)& 1 & 1 & 0.9 \\
 &\,\,\, \cite{open2015estimating} &&& \\\hline
2&Equipoise (maximum uncertainty) & 0.5 & 0.947 & 0.808 \\
&\,\,\,  \cite{freedman1987} &&& \\\hline
3&Consistent with ML-RP & 0.25 & 0.857 & 0.776 \\
&\,\,\, \cite{klein2014} &&& \\\hline
4&Secondary or exploratory analyses & 0.05 & 0.486 & 0.439 \\
\hline
\end{tabular}
\end{table}


\bcx Case 1 of Table  \ref{table.ppv.pred} \ec is \bcx overly optimistic\ec, and represents the values of $\pi$ and $PPV$ assumed by the OSC-RP. \bcx Case 2 \ec  represents maximum experimental uncertainty. In the context of randomized clinical trials (RCT), this has been referred to as \emph{equipoise}, or complete uncertainty regarding the relative efficacy of two experimental treatments \citep{freedman1987} (this will be discussed further in Section \ref{sec.equipoise} below).  For Case 3 we assume $\pi = 0.25$, which is consistent with results reported  in the ML-RP \citep{klein2014}. Case 4 can reasonably model  secondary or exploratory analyses, for which it is anticipated that most hypotheses will be truly  null. 

\bcx
Clearly, the large range of both  $PPV$ and $PPV_{obs}$ found in Table \ref{table.ppv.pred} suggests that the  interpretation of any reported reproducibility rate must refer to a reasonable conjecture regarding the value of $\pi$. 
\ec


\subsection{Prevalence $\pi$ for clinical trials}\label{sec.clinical.trials}

In principle, the value of $\pi$ in clinical trials can be estimated from reported success rates:  
\begin{quote}
Under current US law, drug trials are supposed to be registered on a government website, \texttt{clinicaltrials.gov}. After the study has been completed, researchers are required to post the results within one year -- positive or negative. \citep{hsieh2015}
\end{quote}

The success rate of clinical trials is discussed in the literature, and reported to some degree by research institutes and agencies. The Food and Drug Administration (FDA) reports the success rates of drugs evaluated through the multiple phases of clinical trials.   Approximately 70\% of drugs pass regulatory phase 1, approximately 33\% of drugs pass phase 2, and between 25\% and 30\% of drugs pass phase 3 (see \url{https://www.fda.gov/ForPatients/Approvals/Drugs/ucm405622.htm}). 

\bcx Clinical trials conducted by members of SWOG (formerly the Southwest Oncology Group) were reported positive about 30\% of the time \citep{unger2015, tompa2016}. \ec \bcx Furthermore, \cite{prinz2011believe} \ec  reported a recent fall in success rates for drug development projects, from 28\% to 18\%. Thus, even for some classes of significant primary analyses, some reports estimate effect prevalence $\pi$ to be less than 50\%.

More recently, in \cite{djulbegovic2013} (which we highly recommend to the reader) it is argued that in RCTs $\pi$ has remained close to 50\%  over the last 50 years: ``[o]ur results show that the probability of finding that a new treatment is better than a standard treatment is about 50--60\%, confirming the theoretical predictions we made more than 15 years ago'' \citep{chalmers1997,djulbegovic2007}. 

Of course, it is always possible that the more recent reliance on high-throughput data and other  data-driven methods  has given clinical research a more exploratory character, which could result in a smaller value of $\pi$.

\subsection{Equipoise}\label{sec.equipoise}


As to the question of what $\pi$ \emph{should} be, the case that the ideal value is $\pi = 0.5$ has been consistently made.  It is a basic fact of information theory that the greatest reduction in uncertainty gained by the observation  of a random outcome occurs when outcome probabilities are equal. It could therefore be argued that the optimal use of finite experimental resources occurs when $\pi = 0.5$.  

The concept of ``clinical equipoise'' was introduced in \cite{freedman1987} as ``... a state of genuine uncertainty on the part of the clinical investigator regarding the comparative therapeutic merits of each arm in a trial''.  For a  null hypothesis that an experimental treatment is not better than conventional treatment  this implies $\pi = 0.5$. This becomes an ethical imperative in RCTs if one accepts that a subject should not be assigned a treatment if there is any evidence (statistical or otherwise) that it is inferior to an available alternative. 

The argument that $\pi = 0.5$ is the ideal effect prevalence  can therefore be made from quite distinct points of view.  Possibly this is unattainable, as would be the case for more exploratory studies, such as biomarker or therapeutic discovery based on high-throughput forms of data (although the standard use of multiple testing corrections can control for this). However, this does not argue against such research; rather, it argues for a greater role for validation studies in those cases.

\subsection{Other interpretation of the reproducibility model}\label{sec.other.prevalence.models}

The publication model of Section \ref{sec.model.def}  has been described elsewhere in the literature (see, for example, the exchange in \citep{button2013,hoppe2013, button2013response}). Equation  \eqref{Odds(PPV)2} appears in \cite{ioannidis2005} in the single stage context (under the alarming title ``Why Most Published Research Findings are False'').  However, an additional parameter $u \in [0,1]$ is added to this relation, defined as ``... the proportion of probed analyses that would not have been `research findings,' but nevertheless end up presented and reported as such, because of bias''. When the remaining parameters are fixed, $PPV$ decreases as $u$ increases. Table 4 of \cite{ioannidis2005}  is comparable to Table \ref{table.ppv.pred} in this article, listing sample pairings of the parameters of Equation \eqref{Odds(PPV)2}.  Whether or not $u$ is included in the relation, the conclusion is about the same; that is, for exploratory research, where $\pi$ is probably below some threshold, most published research findings will indeed be false. Where we differ with  \cite{ioannidis2005} is in our claim that low values of $PPV$ and $PPV_{obs}$ can be explained under the assumption that current experimental design and practice are sound. Therefore, if $PPV$ is in some sense too small, one simple corrective is to recognize the role played by validation studies. 

\section{On the use of preliminary data for power analyses}\label{sec.prelim} 

The problem of sample size estimation for a reproducibility study introduces technical issues not normally encountered in conventional experimental design. However, it does bear comparison to the common practice of using preliminary (or pilot) data to estimate the sample size for a future study (we use the terms \emph{preliminary} and \emph{future studies} in this context).  While formulas for sample sizes based on parametric models are routine in statistical practice, their validity may be compromised when parameters have to be estimated. 

This practice takes several forms. The least problematic is the estimation of nuisance parameters such as variance. However, even here the variation induced by the estimation can affect the accuracy of the power analysis considerably. See, for example, \cite{ryan2013} for a summary of the literature on this problem. 

More problematic is the use of preliminary data to estimate the effect size to be powered.  The problem is compounded when that same data is used to determine whether or not to conduct a future study. Here, whether acknowledged or not, the preliminary study is really part of a two-stage decision process, and should be analyzed as such.  

\subsection{A simple power analysis model}\label{sec.power.model} 

We will use the simplest hypothesis test as the representative case. Consider a one-sided test for the mean $\mu$ of a normal distribution with known variance $\sigma^2$. We are given an \textit{iid} sample of size $n$ from distribution $N(\mu,\sigma^2)$ to test null hypothesis $H_o:\mu=0$ against alternative $H_a:\mu>0$. We define standardized effect size $\delta = \mu/\sigma$ and noncentrality parameter ($ncp$) $\eta = \delta\sqrt{n}$.  Most other hypothesis tests have analogous quantities which admit a common interpretation. While the scientific questions concern $\delta$, the problem of power and statistical significance is driven more by $\eta$.     

Let $\bar{X}_{obs}$ be the observed sample mean. Then we have estimates
\beas
\hat{\delta} &=& \frac{\bar{X}_{obs}}{\sigma}, \\
\hat{\eta} &=& \hat{\delta} \sqrt{n} =z_p = z_0 + \eta,
\eeas
where $z_0 \sim N(0,1)$, of which $\phi$, $\Phi$ will denote the density and CDF, respectively. The $P$-value is then the 1-1 transformation $P = 1 - \Phi(z_p)$, obtained from the rejection rule $z_p \geq  z_{\alpha}$.  

Then suppose $n^*$ is fixed as the new sample size for the future study. The $ncp$ for the future study is now $\eta^* = \delta \sqrt{n^*} = \eta \sqrt{n^*/n}$. This gives type II error
\beq
\beta = \Phi\left(z_\alpha - \eta^*\right) = \Phi\left(z_\alpha - \eta \sqrt{\frac{n^*}{n}} \right). \label{eq.beta.1}
\eeq
Equation \eqref{eq.beta.1} can be rewritten to give the standard sample size formula
\beq
R^* = \frac{n^*}{n} = \left(\frac{z_\alpha + z_\beta}{\eta}\right)^2. \label{n.formula.0}
\eeq 
Here, we introduce the notation $R^*$ to emphasize the primary role played by the ratio $n^*/n$, rather than by the sample sizes considered independently. If $z_p$ is accepted as an estimate of $\eta$, an estimate  $\hat{n}^* \approx n^*$ is obtainable by simple substitution:
\bea
\hat{n}^*(z_p,n,\alpha,\beta)  &=& n \left(\frac{z_\alpha+z_\beta}{z_p}\right)^2, \nonumber \\
\hat{R}^*(z_p,\alpha,\beta) &=& \frac{\hat{n}^*(z_p,n,\alpha,\beta) }{n}. \label{n.formula.1}
\eea
Substitution of the ratio $\hat{n}^*/n$ back into \eqref{eq.beta.1} then gives the type II error conditional on $z_p$ as:
\beq
\hat{\beta}(z_p, \eta, \alpha, \beta)  =  \Phi\left(z_\alpha - \eta \frac{z_\alpha + z_\beta}{z_p} \right) 
= \Phi\left(z_\alpha - \frac{z_\alpha + z_\beta}{\frac{z_0}{\eta}  + 1} \right). \label{eq.beta.2} 
\eeq
If we then introduce the crude estimate $z_0/\eta \approx E[z_0/\eta] = 0$ into \eqref{eq.beta.2}, we obtain the approximation $\hat{\beta} \approx \beta$.  

\subsection{One- versus two-sided tests}\label{sec.1vs2}

While the one-sided test offers greater analytical clarity, the two-sided test is commonly used even when a specific effect direction is anticipated. In this case the rejection rule is now $|z_p| \geq z_{\alpha/2}$, so the type II error is
\beq
\beta =  \Phi\left(z_{\alpha/2} - \eta^* \right) - \Phi\left(-z_{\alpha/2}  -  \eta^*  \right). \label{eq.beta.1.2side}
\eeq
Suppose $\eta^*_\beta$ is the solution to Equation \eqref{eq.beta.1.2side} with respect to $\eta^*$. Then set
\beq
R^* = \frac{n^*}{n} \approx \left(\frac{\eta^*_\beta}{\eta} \right)^2. \label{n.formula.0.2side}
\eeq 
A widely accepted approximation gives $\eta^*_\beta \approx z_{\alpha/2} + z_\beta$. Replacing $\eta$ with approximation $z_p$ gives 
\bea
\hat{n}^*(z_p,n,\alpha,\beta)  &=& n \left(\frac{z_{\alpha/2}+z_\beta}{z_p}\right)^2, \nonumber \\
\hat{R}^*(z_p,\alpha,\beta) &=& \frac{\hat{n}^*(z_p,n,\alpha,\beta) }{n}. \label{n.formula.1.2side}
\eea
Substitution of the ratio $\hat{n}^*/n$ back into \eqref{eq.beta.1.2side} then gives conditional type II error 
\beq
\hat{\beta}(z_p, \eta, \alpha, \beta)  =  \Phi\left(z_{\alpha/2}  - \eta \frac{z_{\alpha/2} + z_\beta}{z_p} \right) 
-    \Phi\left( -z_{\alpha/2}  - \eta \frac{z_{\alpha/2} + z_\beta}{z_p} \right).  \label{eq.beta.2.2side} 
\eeq
An argument similar to that following Equation \eqref{eq.beta.2} gives  the approximation $\hat{\beta} \approx \beta$, after noting that the second term in \eqref{eq.beta.2.2side} is close to zero for typical values of $\alpha$.

\subsection{Two-stage power calculation}\label{sec.two.stage}


Whether we use the one- or two-sided test, we are given a conditional type II error $\hat{\beta}(z_p,\eta,\alpha, \beta)$ with the property $\hat{\beta}(\eta,\eta,\alpha, \beta) = \beta$ (with negligible error for the two-sided test), and sample size formula $\hat{n}^*(z_p,n,\alpha, \beta)$. These are given explicitly for each test by Equations \eqref{n.formula.1}, \eqref{eq.beta.2},  \eqref{n.formula.1.2side} and \eqref{eq.beta.2.2side}.  

We then consider two types of decision rule. Both rely on a rejection region   $z_p \in R_{pre}$ applied to the preliminary data. For a one-sided test $R_{pre} = \{ z_p \geq t \}$, for a two-sided test $R_{pre} = \{ |z_p| \geq t \}$. For the  \emph{unconditional decision rule}, we commit to a future study for any $z_p$.  If $z_p \notin R_{pre}$ we  use sample size $\hat{n}^*(t, \alpha, \beta)$, otherwise we use $\hat{n}^*(z_p, \alpha, \beta)$. For the \emph{conditional decision rule}, we only  undertake the future study if $z_p \in R_{pre}$, in which case sample size $\hat{n}^*(z_p, \alpha, \beta)$ is used. 

The actual type II error for each decision rule is then given by the expectations:
\bea
\beta_u(\eta, t, \alpha,\beta) &=& E_\eta\left[ \hat{\beta}(z_p, \eta, \alpha, \beta) I\{ z_p \in R_{pre} \} +   \hat{\beta}(t, \eta, \alpha, \beta) I\{ z_p \notin R_{pre} \}  \right], \nonumber \\
\beta_c(\eta, t,\alpha,\beta) &=& E_\eta\left[ \hat{\beta}(z_p, \eta, \alpha, \beta) \mid z_p \in R_{pre} \right], \label{eq.marginal.beta}
\eea
for the unconditional and conditional rules, respectively, with the corresponding expected sample size ratios calculated similarly:
\bea
R^*_u(\eta,t, \alpha,\beta) &=& E_\eta\left[ \hat{R}^*(z_p,\alpha,\beta) I\{ z_p \in R_{pre} \} +   \hat{R}^*(t, \alpha, \beta) I\{ z_p \notin R_{pre} \}  \right], \nonumber \\
R^*_c(\eta,t, \alpha,\beta) &=& E_\eta\left[ \hat{R}^*(z_p,\alpha,\beta)  \mid z_p \in R_{pre} \right].  \label{eq.marginal.rstar}
\eea
The value $\beta$ in the argument of $\beta_u(\eta, t,\alpha,\beta)$ or $\beta_c(\eta, t,\alpha,\beta)$ is the nominal type II error, the latter quantities being the actual, or realized, type II error.

\begin{figure}[h] 
\includegraphics[width=6in,height=5.25in,viewport=20 50 550 480, clip]{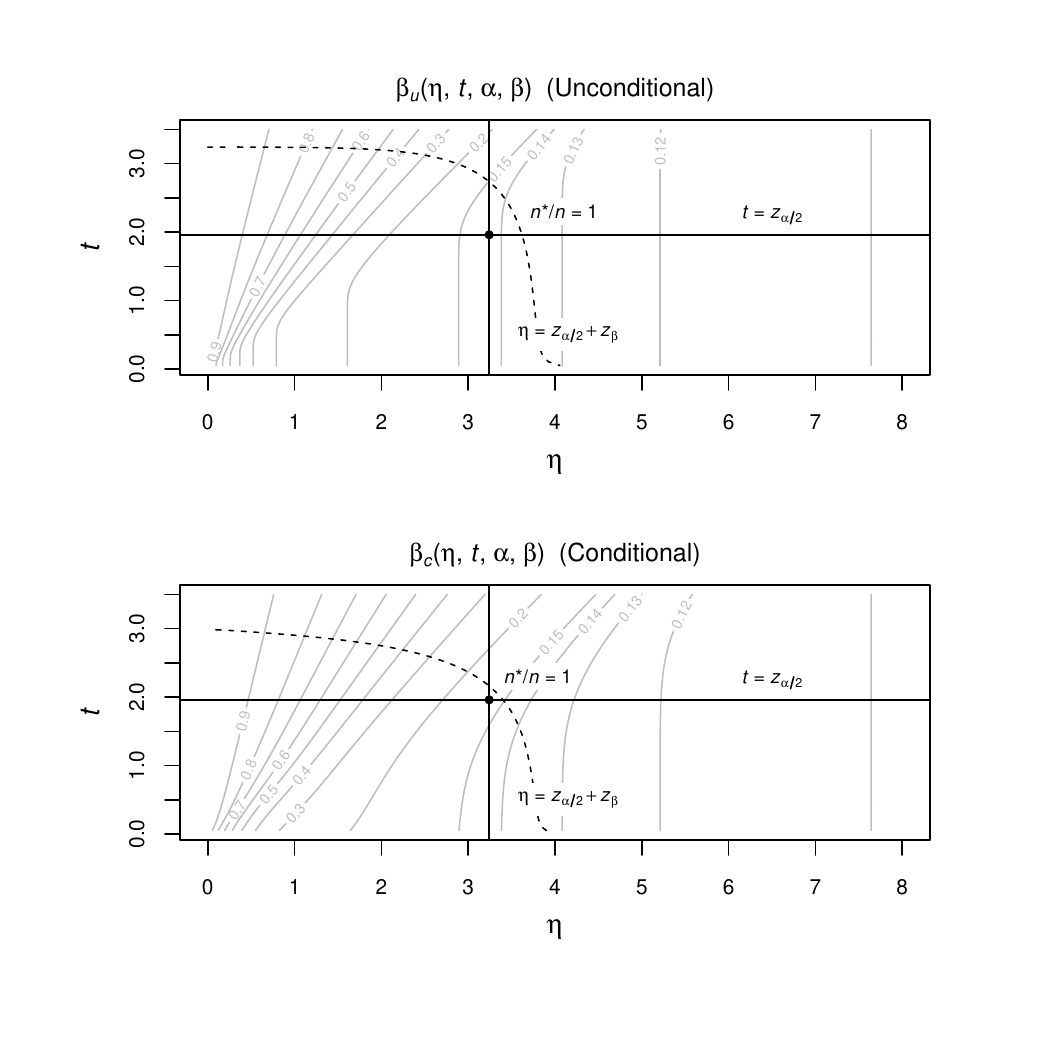}
\caption{Contour plots of $\beta_u(\eta,t,\alpha, \beta)$ and  $\beta_c(\eta,t,\alpha, \beta)$ for fixed nominal type I, II errors $\alpha = 0.05$, $\beta = 0.1$. Contour for $n^*/n = 1$ is superimposed using a dashed line [- - -]. Values $t = z_{\alpha/2}$ and $\eta = z_{\alpha/2} + z_\beta$ are indicated for convenience.}\label{fig-beta}  
\end{figure}

\begin{figure}[h] 
\includegraphics[width=6in,height=5.25in,viewport=20 50 550 480, clip]{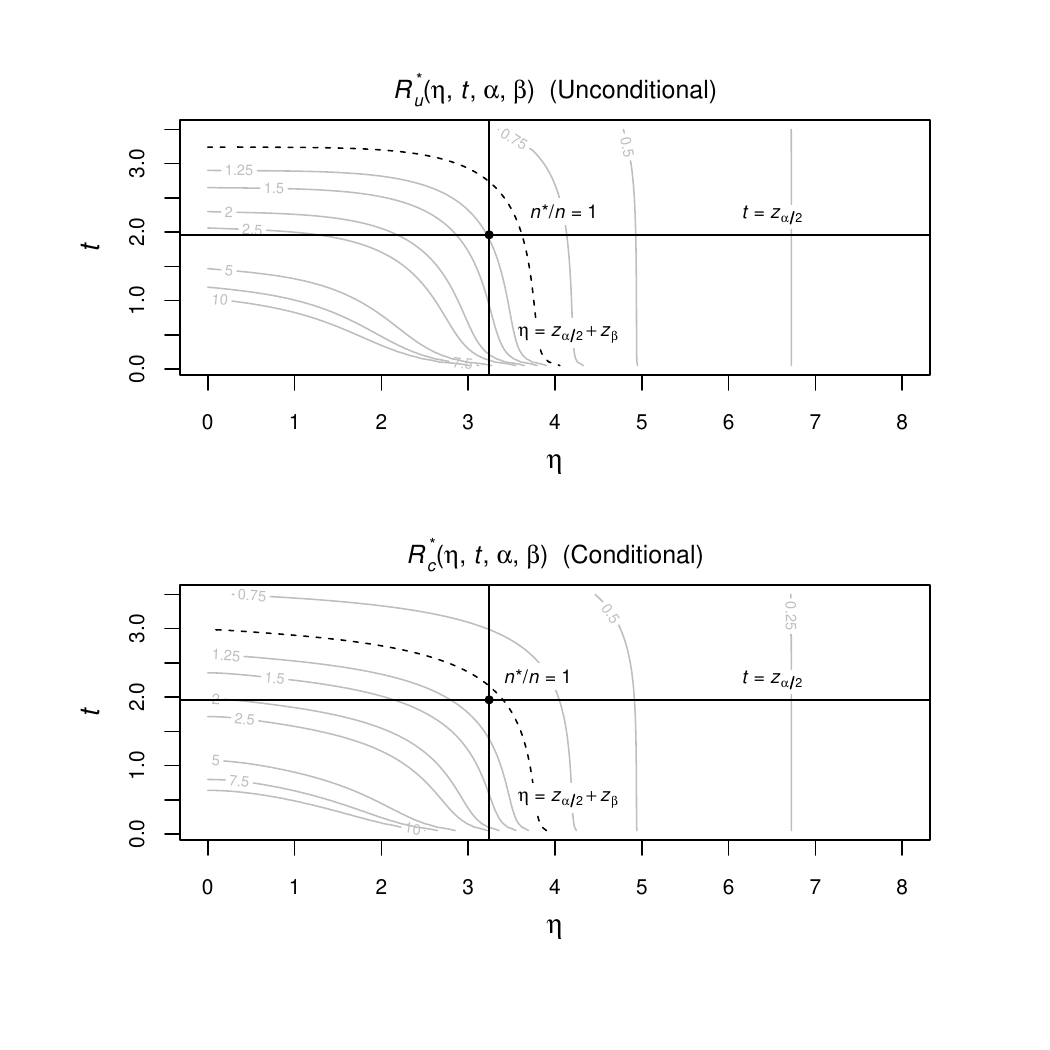}
\caption{Contour plots of $R^*_u(\eta,t,\alpha, \beta)$ and  $R^*_c(\eta,t,\alpha, \beta)$ for fixed nominal type I, II errors $\alpha = 0.05$, $\beta = 0.1$. Contour for $n^*/n = 1$ is indicated by a dashed line  [- - -]. Values $t = z_{\alpha/2}$ and $\eta = z_{\alpha/2} + z_\beta$ are indicated for convenience.}\label{fig-nstar}  
\end{figure}

For the two-sided test, with  nominal type I, II errors $\alpha = 0.05$, $\beta = 0.1$, contour plots of $\beta_u(\eta,t,\alpha, \beta)$ and  $\beta_c(\eta,t,\alpha, \beta)$ are given in Figure \ref{fig-beta}, while contour plots of $R^*_u(\eta,t,\alpha, \beta)$ and  $R^*_c(\eta,t,\alpha, \beta)$ are given in Figure \ref{fig-nstar}.   All contour plots contain alternative axes  $t = z_{\alpha/2}$ and $\eta = z_{\alpha/2} + z_{\beta}$. Their intersection represents  a typical power analysis scenario.  A future study is planned if $|z_p| \geq t = z_{\alpha/2}$, while the value $\eta = z_{\alpha/2} + z_{\beta}$ gives a nominal power of $1-\beta$, as is typically intended. 

In addition, the contour lines $R^*_u(\eta,t,\alpha, \beta) = 1$ and  $R^*_c(\eta,t,\alpha, \beta) = 1$ are superimposed on the plots for $\beta_u(\eta,t,\alpha, \beta)$ and  $\beta_c(\eta,t,\alpha, \beta)$ (Figure \ref{fig-beta}). Essentially, this contour represents approximately the case $n^*/n = 1$.

Perhaps the most striking feature of Figure \ref{fig-beta} is the degree to which the actual type II error exceeds the nominal value of $\beta = 0.1$. While this discrepancy is moderate for $\eta \geq z_{\alpha/2} + z_{\beta}$ and $t \leq z_{\alpha/2}$, bounded above by about $\beta = 0.15$, it is quite severe for smaller values of $\eta$. For example, for $\eta = 1$ and $t =  z_{\alpha/2}$ we have $\beta_c(\eta,t,\alpha, \beta) \approx 0.7$. Also, for the same threshold $t =  z_{\alpha/2}$ the larger effect size $\eta = z_{0.025} = 1.96$ yields actual type II error  $\beta_c(\eta,t,\alpha, \beta) \approx 0.3$.  

Furthermore,  this high discrepancy region corresponds roughly with the region $n^*/n > 1$.  The rather troubling implication of this is that whenever preliminary data is used to power a future study in this way, if it happens that the future sample size is larger than the preliminary sample size, this means that the actual type II error may be significantly larger than the nominal value. Of course, such an increase in sample size between preliminary and future studies is almost always anticipated.  

Clearly, if we use the nominal value $\beta = 0.1$, the power will be overestimated, since the actual type II error will be $\beta_c(\eta,t,\alpha, \beta) > \beta$. Of course, a valid power analysis is possible, as long as the decision theoretic character of the problem is acknowledged, and several approaches exist. The next example illustrates a strategy of accepting the nominal type II error probability, then verifying that any discrepancy is not too great. \\

\noindent\textbf{Example 1.} Suppose we are interested in detecting a clinically significant effect size $\delta = \mu/\sigma \geq 1/2$. From Figure \ref{fig-beta} we can see that a value of $\eta = 4$ results in only modest discrepancy. Suppose we decide to proceed with a future study if $|z_p| \geq t = z_{0.025}$ (i.e. the conditional decision rule). The sample size is obtained from
\bdm
\eta = \delta\sqrt{n}  =  \frac{\sqrt{n}}{2} = 4, 
\edm
or $n = 64$.  From Figure \ref{fig-beta} (bottom plot) a nominal type II error $\beta = 0.1$ results in an actual value of $\beta \approx 0.135$. This may be regarded as an acceptable approximation. However, from Figure \ref{fig-nstar}, it can be seen that $E[\hat{n}^*/n] < 1$, so that the future study is likely to have a \emph{smaller} sample size than the preliminary study.  
\qed \\

In Example 1, the strategy was to select preliminary sample size $n$ to be large enough to ensure a useful  estimate of the effect size for a power analysis. However, this forces $E[\hat{n}^*/n] < 1$. In effect, this strategy turns the preliminary study into the main study.  

Suppose instead we conduct a power analysis for the two-stage decision process, attaining a target type II error probability by anticipating, then adjusting for, discrepancy  in the power estimate.  The advantage of this approach is that an accurate power estimate does not depend on an accurate estimate of effect size $\eta$. This is illustrated in the next example.  \\

\begin{figure}[h] 
\includegraphics[width=6in,height=5.25in,viewport=20 50 550 480, clip]{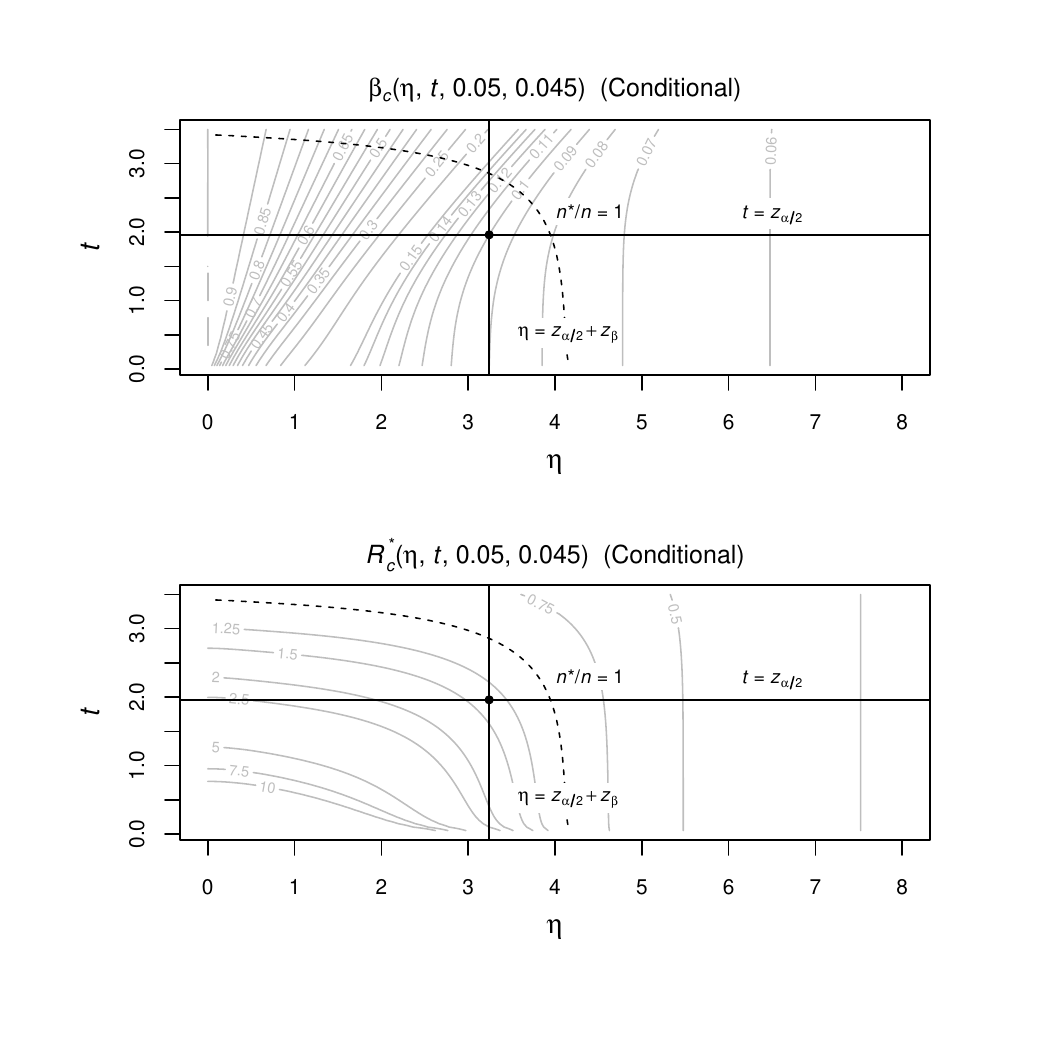}
\caption{Contour plots of $\beta_c(\eta,t,0.05, 0.05)$ and $R^*_c(\eta,t,0.05, 0.05)$. Contour for $n^*/n = 1$ is indicated by a dashed line  [- - -]. Values $t = z_{\alpha/2}$ and $\eta = z_{\alpha/2} + z_\beta$ are indicated for convenience.}\label{fig-example}  
\end{figure}

\noindent\textbf{Example 2.} Suppose we wish to attain $\beta = 0.1$ for a future study, again to be carried out only if  $|z_p| \geq t = z_{0.025}$.  We may simply adjust the nominal value $\beta$  downwards. To see this, consider Figure \ref{fig-example}, which shows contour plots of $\beta_c(\eta,t, 0.05, 0.045)$ and $R^*_c(\eta,t,0.05, 0.045)$. We are adjusting the nominal type II error to $\beta =0.045$. Note that we now have value of $\beta_c \approx 0.1$ at the point defined by $t = z_{0.025}$ and $\eta = z_{0.025} + z_{0.045}$.  

As for Example 1,  suppose we are interested in a clinically significant effect size $\delta = \mu/\sigma \geq 1/2$. We then plan for 
$$
\eta = \delta\sqrt{n}  =  \frac{\sqrt{n}}{2} =  z_{0.025} + z_{0.045} \approx 3.66,
$$
so we need a preliminary sample size $n \geq (2 \times 3.66)^2 = 53.6$, or $n = 54$. From Figure \ref{fig-example} we can see that the expected value of the ratio $\hat{n}^*/n$ is in the interval $[1.25,1.5]$. Given the threshold $t = z_{0.025}$, the ratio could be as high as
$$
\hat{n}^*/n  = \left( \frac{z_{\alpha/2} + z_\beta}{t} \right)^2 \approx \left(\frac{3.66}{1.96}\right)^2 = 3.49.
$$
We therefore expect a larger sample size for the future study. Clearly, this example better approaches the expectation of a power analysis based on preliminary data than does Example 1.
\qed

\subsection{Estimation of effect size}\label{sec.actual.effect.size}

 We next consider the problem of estimating $\eta$ given $z_p$. This is necessarily a challenging problem, since not only is it based on a single sample $z_p \sim N(\eta, 1)$, but $z_p$ is also truncated. 

Here, it serves our purpose to assume that $z_p$ is positive so that only values $z_p \geq t$ are observed. Here,  $t$ may be $z_\alpha$ or $z_{\alpha/2}$ according to the type of test.  We may then use $z_p$ to calculate the maximum likelihood estimate of $\eta$ using the truncated density
$$
f_{z_p}(y; \eta) = \frac{\phi(y - \eta)}{1 - \Phi(t - \eta)} I\{ y \geq t\}.
$$
It is easily verified that as $\eta \rightarrow -\infty$ the density $f_{z_p}(y; \eta)$ is concentrated on an increasingly small interval  with left endpoint $t$, approaching a point mass of 1 at $t$. Thus, the MLE $\hat{\eta}_{MLE}$ exists for all $z_p > t$, with   $\lim_{z_p \downarrow t} \hat{\eta}_{MLE} = -\infty$. We may then  reasonably define  $\hat{\eta}_{MLE} = -\infty$ when $z_p = t$. 
 
\begin{figure} 
\centering
\includegraphics[width=3in,height=3in,viewport=0 20 450 450, clip]{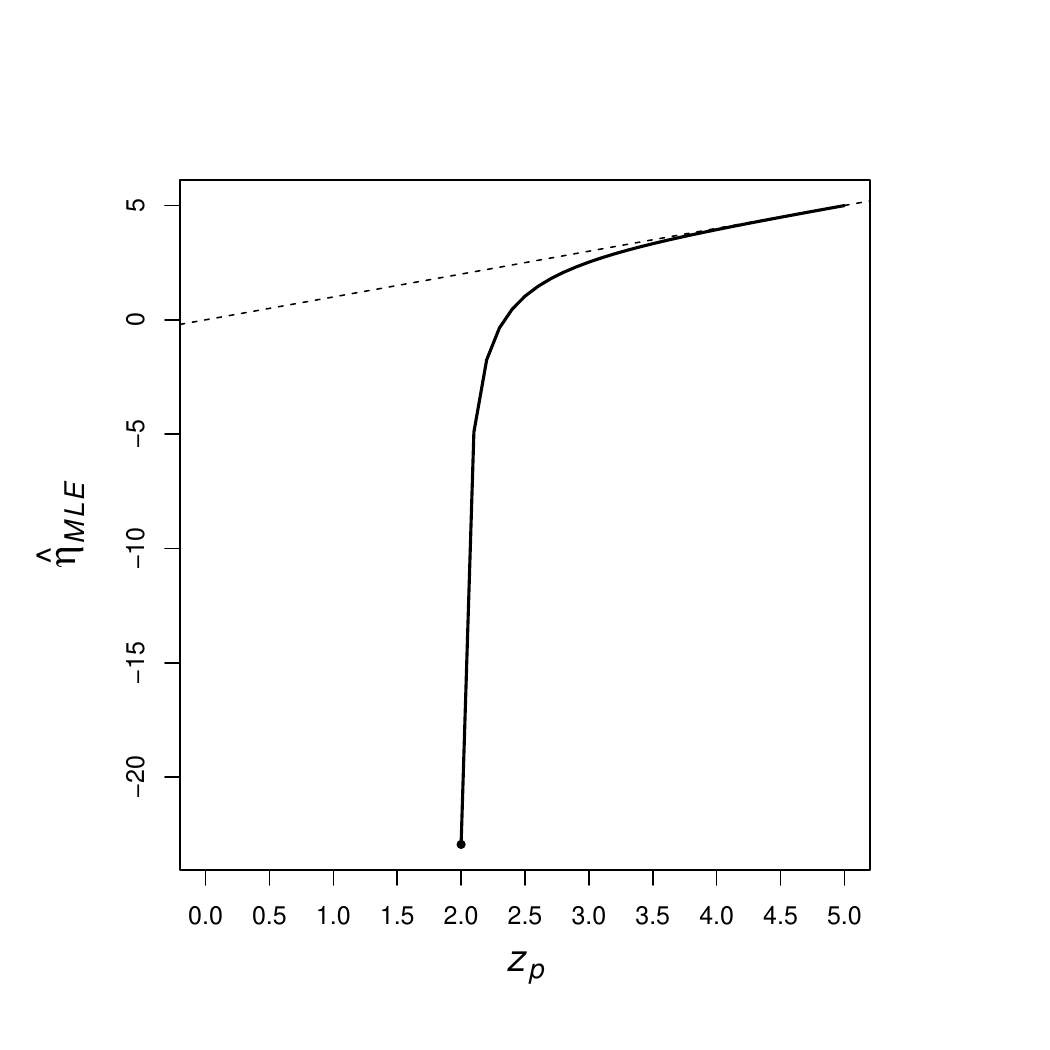}
\caption{Maximum likelihood estimate of $\eta$ based on a single observation of $z_p \in [2.0,5.0]$, truncated at $z_p \geq t = z_{0.025}$. The identity is indicated by a dashed line  [- - -].}\label{fig-etamle}
\end{figure}

\begin{figure} 
\centering
\includegraphics[width=4in,height=4in,viewport= 40 50 440 450, clip]{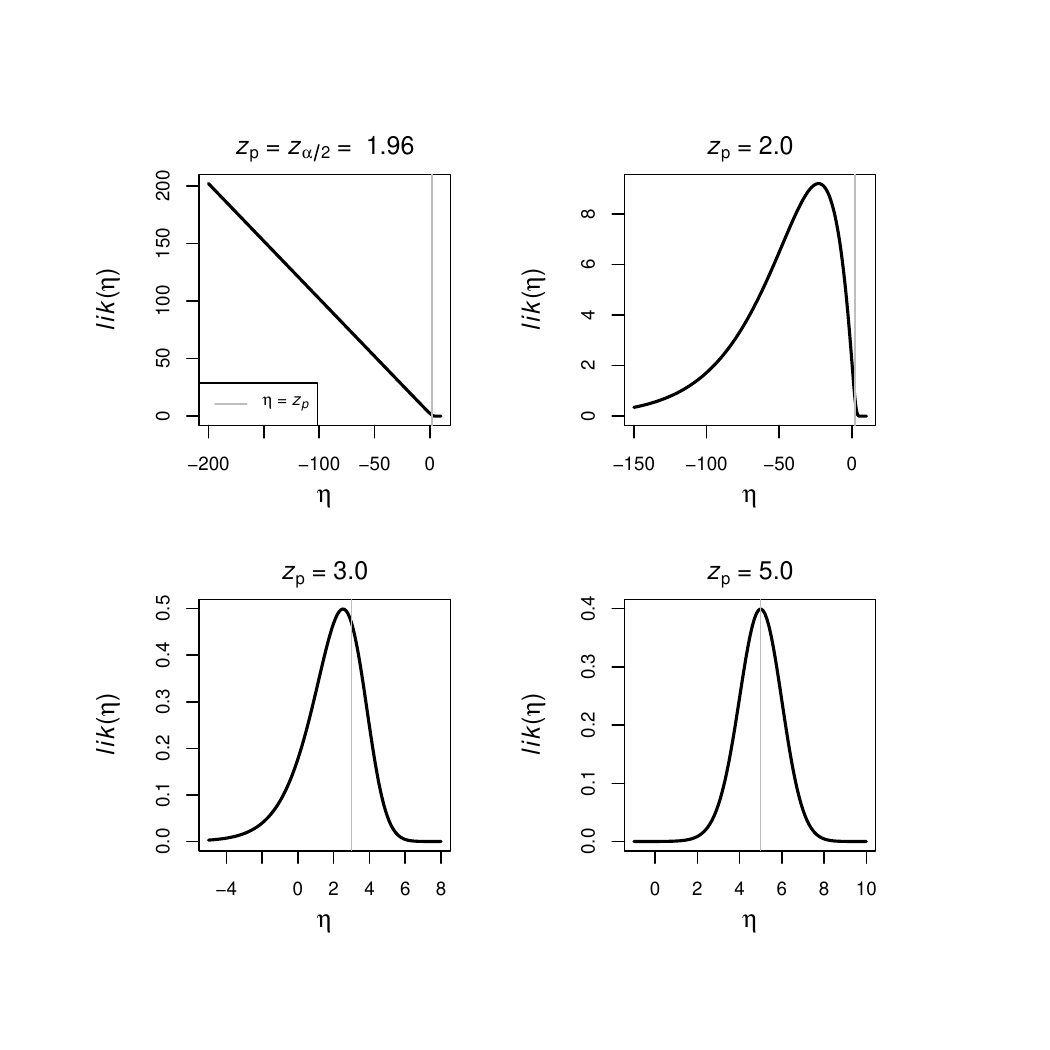}
\caption{Likelihood functions of $\eta$ based  on a single observation of $z_p = z_{0.025}$, 2.0, 3.0, 5.0, truncated at $z_p \geq t = z_{0.025}$.  The value $\eta = z_p$ is indicated by the gray line.}\label{fig-etamle2}
\end{figure}

The deterministic relationship between $z_p$ and $\hat{\eta}_{MLE}$ is shown in Figure \ref{fig-etamle}, using $t = z_{0.025}$. The identity is indicated by a dashed line. The plot only shows values $z_p \geq 2.0$ noting that  $\hat{\eta}_{MLE}$ is unbounded from below as $z_p \downarrow z_{0.025}$. The approximation $\eta \approx z_p$ appears reasonable for, say, $z_p \geq 3.0$, but at smaller values there seems to be little relationship between the estimator and $\eta$. Figure \ref{fig-etamle2} shows the likelihood function for selected values $z_p = z_{0.025}$, 2.0, 3.0, 5.0.  As discussed above, for $z_p  = z_{0.025}$ we set $\hat{\eta}_{MLE} = -\infty$, and, accordingly, the likelihood function increases indefinitely as $\eta \rightarrow -\infty$. For the remaining values $z_p = $ 2.0, 3.0, 5.0  we have well-defined estimates $\hat{\eta}_{MLE} \approx$  -22.938, 2.52, 4.996, respectively. The point estimate for $z_p =  2.0$ is not reasonable, and the inference is simply that the plausible range of values for $\eta$ extends from a value slightly larger than 2.0 to essentially arbitrarily small negative values.  The range of plausible values for $z_p =$  3.0, 5.0 is more useful, although impractically wide for a formal estimate.  Nonetheless, for $z_p = 5.0$ an inference that $\eta > z_{0.025}$ is apparently supported. 

Clearly, any procedure that depends on an accurate estimate of $\eta$ cannot be recommended. However, 
as shown in  Example 2,  a valid and useful power analysis is possible that does not rely on such an estimate.

\subsection{The OSC-RP revisited}\label{sec.rprev}

The issues underlying effect size estimation described here are directly relevant to the protocol used for the OSC-RP, which uses the same direct effect size estimate, equivalent to $\eta \approx z_p$, as described in the  supplementary material of \cite{open2015estimating}: ``After identifying the key effect, power analyses estimated the sample sizes needed to achieve 80\%, 90\%, and 95\% power \textbf{to detect the originally reported effect size}'' (emphasis added).  Conditional truncation (Section \ref{sec.two.stage}) is also implicit in the OSC-RP protocol, equivalent to $|z_p| \geq  z_{0.025}$ in our own model for a two-sided test. 

In addition, the supplementary material adds: ``[p]ost-hoc calculations showed an average of 92\% power \textbf{to detect an effect size equivalent to the original studies}'' (emphasis added). 

Data on the studies is available as supplementary material for \cite{open2015estimating}. The master file contains 167 records (some original papers are represented more than once). For 100 of these, original and replication data are classified as complete, so that this subset is used for the replication assessment.  For each of these, a significance indicator ($P \leq 0.05$) is included for the original and replicated study (Table \ref{table.rp.endpoint}). In fact, three  of the original studies are reported as nonsignificant, therefore we base our own calculations on the remaining 97. From the study we use the quantities listed in Table \ref{table.rp.variables}. The sample size $N$ is the number of nonmissing values of the subset of 97 originally significant studies.   A number of measures of reproducibility are reported in \cite{open2015estimating}, but we consider specifically the rule $P \leq 0.05$. 

\begin{table}
\centering
\caption{Paired outcome frequencies  (Nonsignificant/Significant) for the $N = 100$ observed and replicated studies used in \cite{open2015estimating} to assess reproducibility. Significance is defined as $P \leq 0.05$. }\label{table.rp.endpoint} 
\begin{tabular}{ll|rr}\hline
&& \multicolumn{2}{c}{Replication} \\
&&Nonsignificant&Significant \\ \hline
Original & Nonsignificant & 2 & 1\\
 & Significant & 62 & 35 
 \end{tabular}
\end{table}

\begin{table}
\centering
\caption{Data from OSC-RP used in calculations of Section \ref{sec.rprev}. Columns refer to spreadsheet file \ttt{rpp\_data.csv} (available as supplementary material to \cite{open2015estimating} at  \url{https://osf.io/fgjvw/}. Sample size refers to the subset of 97 studies classified as complete and originally significant.}\label{table.rp.variables}  
\begin{tabular}{l|lcr}\hline
Description & Column Header & Column & $N$ \\ \hline
Sample size of original study & \ttt{N (O)} & \ttt{AZ} & 97\\
Sample size of replication study & \ttt{N (R)} & \ttt{BR} & 97\\
Type of effect & \ttt{Type of effect (O)} & \ttt{BE} & 97\\
Replication power & \ttt{Power (R)} & \ttt{BY} & 94 \\
$P$-value of original study & \ttt{T\_pval\_USE..O.} & \ttt{DH} & 96 \\
Original study significant (=1) & \ttt{T\_sign\_O} & \ttt{EA} & 97 \\ 
Replication study significant (=1) & \ttt{T\_sign\_R} & \ttt{EB} & 97 
\end{tabular}
\end{table}

\begin{figure}[h] 
\includegraphics[width=5in,height=4in,viewport=40 20 500 435, clip]{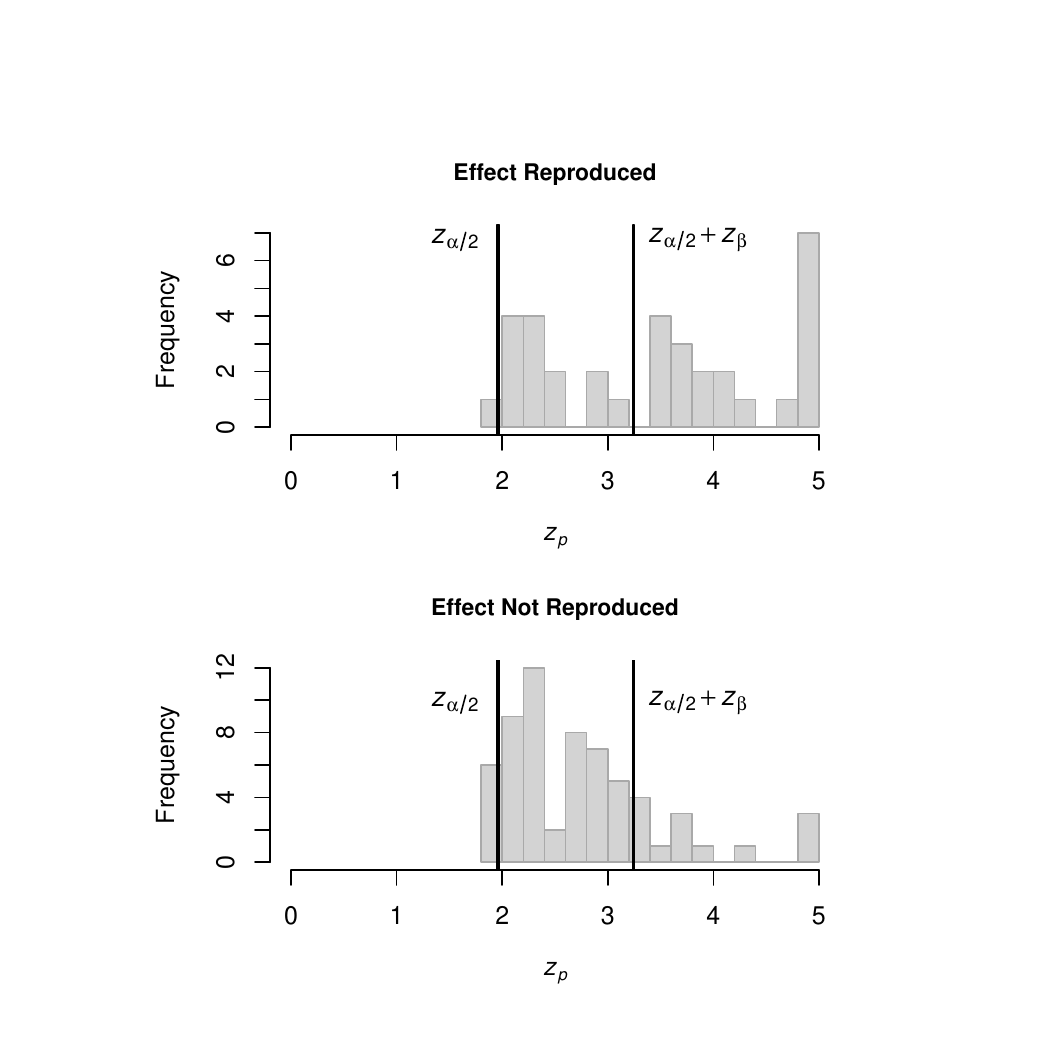}
\caption{Histograms of imputed values of $z_p$ by effect reproduction outcome.  The values of $z_{\alpha/2}$ and $z_{\alpha/2} + z_{\beta}$, $\alpha = 0.05$, $\beta = 0.1$ are superimposed for reference. The distributions differ significantly ($P = 0.0016$, Wilcoxon rank sum test).}\label{fig-zhist}
\end{figure}

The type of hypothesis test varies considerably. We will therefore apply our power model by analogy. We assume that all relevant tests used in the OSC-RP are identical to the two-sided test described in Sections \ref{sec.power.model}-\ref{sec.1vs2}, which then yielded the reported $P$-values referred to in Table \ref{table.rp.variables}. Observed effect sizes can be easily imputed by the equation $z_p = \Phi^{-1}(1-P/2)$. A number of $P$-values were given as 0, or were too small to be justified by standard normal approximation theory (e.g. $P = 1.39\times10^{-43}$), so a lower bound of $P \geq 10^{-6}$ was imposed (representing about 4.9 standard deviations). This affected ten $P$-values.   A histogram of these imputed values  is shown in Figure \ref{fig-zhist} separately for effect reproduction outcome (Table \ref{table.rp.endpoint}). The distributions differ significantly ($P = 0.0016$, Wilcoxon rank sum test). The values of $z_{\alpha/2}$ and $z_{\alpha/2} + z_{\beta}$, $\alpha = 0.05$, $\beta = 0.1$ are superimposed for reference.  The values are bounded below by $z_{\alpha/2}$ as expected. Interestingly, for the studies with a positive reproduction outcome the values of $z_p$ seem roughly uniformly distributed within the available range, while for the studies with a negative reproduction outcome the values of $z_p$ appear to form a tail of a truncated distribution, suggesting that a higher proportion of these are from the upper tail of a normal distribution with mean $\eta$ well below $z_{\alpha/2}$, as would be predicted by our reproducibility model.

 Our next step  is to estimate the actual power attained by the OSC-RP using the conditional decision process described in Section \ref{sec.two.stage}.  This is appropriate given the selection of studies for which $|z_p| \geq z_{\alpha/2} = t$.  Then for each study the actual type II error is the quantity $\beta_c(\eta,t,\alpha, \beta)$. We fix $\alpha = 0.05$, $t = z_{\alpha/2}$. The nominal value of $\beta$ is set separately for each study, using the replication power listed in Table \ref{table.rp.variables}.   
 
This requires knowledge of $\eta$, which obviously varies across the studies. In principle, $\eta$ may be estimated by $z_p$, as is done in the OSC-RP protocol. However, in Section \ref{sec.actual.effect.size} it was shown that $z_p$ may significantly overestimate $\eta$  (in fact, it enforces the quite unreasonable assumption that $\eta \geq z_{\alpha/2}$). We use instead the approximation $\eta \approx \hat{\eta}_{MLE}$. We additionally assume that extremely small estimates of $\eta$ are interpretable only as being negative, so the estimates $\hat{\eta}_{MLE}$ will be bounded below by 0.

This leaves $N = 93$ samples for which both replication power and $P$-value (and therefore imputed $\eta$) are available (Table \ref{table.rp.variables}).  After these calculations are applied, the mean type II error is now $\bar{\beta}^* = 0.468$, considerably higher than the average $0.08$ reported by the OSC-RP.

We first note that an observed reproducibility rate of 35/97 (Table \ref{table.rp.endpoint}) yields a 95\% confidence interval of  $PPV_{obs} \in (0.266,0.465)$ (Clopper-Pearson). 

Next, consider the appropriate value of effect prevalence $\pi$. As pointed out above, the OSC-RP essentially assumes $\pi = 1$, which is unrealistically high.  Suppose, in fact, $\pi = 0.25$ (Case 3 of  Table \ref{table.ppv.pred}) and we accept $PPV = 0.857$ (we argue below that the rates reported by the ML-RP are compatible with these values).  If  we use the average $\bar{\beta}^* = 0.08$  reported by the OSC-RP and $\alpha^* = 0.05$, this yields 
$$
PPV_{obs} = PPV (1 - \beta^*) + (1-PPV) \alpha^* = 0.796,
$$
which is well above the rate reported by the OSC-RP, and not statistically compatible with it.  However, using the value $\bar{\beta}^* = 0.468$ yields $PPV_{obs} = 0.463$, just below the upper bound of our confidence interval.

However, it is possible to say something more about what $\pi$ may be for the OSC-RP population. From the report (emphasis added):
\begin{quote}
\textbf{By default, the last experiment reported in each article was the subject of replication.} This decision established an objective standard for study selection within an article and was based on the intuition that the first study in a multiple-study article (the obvious alternative selection strategy) was more frequently a preliminary demonstration ...  For the purposes of aggregating results across studies to estimate reproducibility, a key result from the selected experiment was identified as the focus of replication. The key result had to be represented as a single statistical inference test or an effect size. \citep{open2015estimating}
\end{quote}
Reasonably, the intention appears to be to confine the OSC-RP to primary analyses which are more likely to be adequately powered. However, the OSC-RP categorizes the \emph{type of effect} (Table  \ref{table.rp.variables}). Most effect types are \ttt{main effect} ($n = 49$) or \ttt{interaction} ($n = 37$). In   \cite{open2015estimating} it is noted that, ``[f]or more complex designs, such as multivariate interaction effects, the quantitative analysis may not provide a simple interpretation ...''. 

Table  \ref{tab.effect.type}  gives the observed reproducibility rate $PPV_{obs}$ for each effect type. For main effects and interactions respectively,   the values of $PPV_{obs}$ were 23/49 = 46.9\%  and 8/37 = 21.6\%, which are significantly different ($P = 0.023$, Fisher's exact test).  
 
This means that among main effects, the observed reproducibility rate is very close to that predicted by our model for an effect prevalence $\pi = 0.25$ (0.463 predicted, 0.469 observed). 

Suppose we next conjecture an effect prevalence of $\pi = 0.05$ (Case 4 of Table  \ref{table.ppv.pred}). Under our assumptions, this leads to $PPV = 0.486$, and with an actual type II error of  $\bar{\beta}^* = 0.468$ a value of $PPV_{obs} = 0.284$ is predicted. This is  reasonably close to the observed  rate $8/37 = 0.216$ among interaction effects, noting that a 95\% confidence interval is given by $PPV_{obs}  \in (0.098, 0.382)$ (Clopper-Pearson). 

Thus, the difference in reproducibility rate by effect type is statistically significant, and  can be explained by differences in $\pi$.  This  is reasonable to expect given that the set of all interactions (in the OSC-RP largely associated with ANOVA models) will be larger than the set of all main effects.   

To summarize, the apparently low reproducibility rate reported by the OSC-RP can be predicted using standard statistical methodology.

\begin{table}[ht]
\centering
\caption{Effect reproduction rates by effect type (see Table  \ref{table.rp.variables}).}\label{tab.effect.type}  
\begin{tabular}{l|rr|r|r}
  \hline
  & \multicolumn{2}{c|}{Reproduced} & & \\
Effect Type  & No & Yes & Total & $PPV_{obs}$ \\    \hline
 \ttt{binomial test} & 0 & 1 & 1 & 1.000 \\ 
  \ttt{contrast} & 1 & 0 & 1 & 0.000 \\ 
  \ttt{correlation} & 4 & 2 & 6 & 0.333 \\ 
  \ttt{focused interaction contrast} & 1 & 0 & 1 & 0.000 \\ 
  \ttt{interaction} & 29 & 8 & 37 & 0.216 \\ 
  \ttt{main effect} & 26 & 23 & 49 & 0.469 \\ 
  \ttt{regression} & 0 & 1 & 1 & 1.000 \\ 
  \ttt{trend} & 1 & 0 & 1 & 0.000 \\ 
   \hline
\end{tabular}
\end{table}

\subsection{Other reproducibility projects}\label{sec.other.rp}

For the ML-RP study  reported in  \cite{klein2014}, effects from 13 studies in the field of psychology were replicated. The main difference with the protocol used by the OSC-RP is that 35--36 replications were used for each study. The effective value of $\alpha^*$ and $\beta^*$ approaches $0$ as the number of replications increases, and thus for a large number of replications we have $PPV \approx PPV_{obs}$, by Equation \eqref{PPVobs}. Therefore, for the ML-RP the reported value $PPV_{obs} = 11/13 = 84.6\%$ is essentially a direct estimate of $PPV$, which is very close to the predicted value  $PPV = 0.857$  given an effect prevalence of $\pi = 0.25$ (Case 3 of Table \ref{table.ppv.pred}). 


The replication study  ECO-RP reported in \cite{camerer2016}, involving  18 experimental studies  in the field of economics, used essentially the same protocol as OSC-RP.  A reported $11/18 = 61.1\%$ of these studies resulted in significant replications.  The sample size formula Equation \eqref{n.formula.1.2side} is given explicitly in the supplemental material  of \cite{camerer2016}, so that empirically observed effect sizes are used directly. Nominal type I, II errors used were $\alpha = 0.05$, $\beta = 0.1$. We estimated the actual type II error $\bar{\beta}^*$ using the procedure of  Section \ref{sec.rprev}.  The $P$-values used by the ECO-RP are given in Table S1 of \cite{camerer2016}, but were supplemented by the individual reproducibility reports when given only as $P<0.001$ (\url{experimentaleconreplications.com}). In addition, the same lower bound  $P \geq 10^{-6}$ used in the analysis of the OSC-RP data was imposed (Section \ref{sec.rprev}). Two $P$-values of the original studies were larger than 0.05 ($P = 0.057, 0.07$). For one of those studies significance was defined as $P \leq 0.1$. For convenience, we used this as a threshold, that is  $t = z_{0.05}$, although $\alpha = 0.05$ remains the nominal type I error for the replication studies.  For frequency $11/18$  a 95\% confidence interval is given by $PPV_{obs} \in (0.357, 0.827)$ (Clopper-Pearson). 

Using the method applied above,  the estimated average  of the actual type II error  obtained was  $\hat{\beta}^* = 0.441$, quite similar to that value of $0.468$ obtained for the OSC-RP data. For a prevalence of $\pi = 0.25$ with $PPV = 0.857$ (Case 3 of Table \ref{table.ppv.pred}), the predicted observed reproducibility rate is $PPV_{obs} = 0.486$, and therefore is compatible with the observed rate (for comparison, if those two $P$-values exceeding 0.05 are deleted and the threshold parameter increased to $t = z_{0.025}$, we obtain $\hat{\beta}^* = 0.499$ with predicted $PPV_{obs} = 0.437$).  

\section{Prevailing views on effect size estimation and power analyses}\label{sec.prevailing.views}


Much of our discussion has focused on the possibility  that the replication protocol used by the OSC-RP may lead to over-estimation of effect size, and therefore under-estimation of power for its replication studies. In fact, this possibility is acknowedged by the OSC-RP itself, in the supplementary material  of \cite{open2015estimating}:  ``[n]ote that these power estimates do not account for the possibility that the published effect sizes are overestimated because of publication biases. Indeed, this is one of the potential challenges for reproducibility''. 

It is therefore important to review what has already been acknowledged in the literature regarding this issue. In fact, the problem with the use of preliminary data for the  estimation of  effect size, and as part of a decision rule, has been widely acknowledged  \citep{lane1978, kraemer2006, leon2011, westlund2016}, and these reports are consistent with the findings of Section \ref{sec.prelim}.   Furthermore, that this problem might exist is quite intuitive. As noted in \cite{lane1978}: ``[e]xperiments that find larger differences between groups than actually exist in the population are more likely to pass stringent tests of significance and be published than experiments that find smaller differences. Published measures of the magnitude of experimental effects will therefore tend to overestimate those effects''. 

 Avoiding this procedure is recognized as good statistical practice. For example, ``[s]ince any effect size estimated from a pilot study is unstable, it does not provide a useful estimation for power calculations ...'', from \emph{Pilot Studies: Common Uses and Misuses},  National Center for Complimentary and Integrative Health (\url{ https://nccih.nih.gov/grants/whatnccihfunds/pilot_studies}). 
 

\subsection{\emph{A priori} effect size}

An obvious alternative to estimation is to base effect size on cost,  benefit or clinical relevance, in which case estimation is not needed.  A good example is the well-known convention proposed in \cite{cohen1988}  that for a two sample treatment effect, $d = 0.2$, $0.5$, $0.8$ be considered small, medium or large, where $d = |\mu_1 - \mu_2|/\sigma$.  Also see  \cite{keefe2013} for an interesting discussion on effect size determined by cost or benefit. 


\subsection{Ethical aspects} 
 
In Section \ref{sec.equipoise} we reviewed a number of arguments concluding that the ideal effect prevalence was $\pi = 0.5$. These had in common the assumption that the purpose of an experiment is to reduce uncertainty (equivalently, gain knowledge). It could therefore be argued that this issue has an ethical dimension.  Suppose we accept that conducting an experiment is unethical, or at least wasteful, if the outcome is already known. The issue then extends to the use of preliminary data to estimate an effect. After all, adding the qualifier ``within statistical error'' to the phrase ``outcome is already known'' doesn't  alter the dilemma in any important way. 

That this is true for an RCT  is clear, since the consequence is to administer treatments to some study subjects that the investigators believe to be inferior, if only in a probabilistic sense. This  logic extends to any branch of experimental science, if we accept the need to allocate finite resources in as efficient a manner as possible. As stated in \cite{kraemer2006}: 
\begin{quote}
... [a]t the time of the design of an RCT, the true effect size is unknown and cannot be used to define the critical value at which power is computed. Indeed, if the true effect size is known a priori with enough confidence to be used to calculate the necessary sample size, conducting an RCT is clinically unethical. 
\end{quote}
Under this point of view, the role of a preliminary study is  to test the feasibility of a future study. Power analyses can then be based on \emph{a priori} effect sizes. From \emph{Pilot Studies: Common Uses and Misuses},  National Center for Complimentary and Integrative Health (\url{ https://nccih.nih.gov/grants/whatnccihfunds/pilot_studies}):
\begin{quote}
Pilot studies should not be used to test hypotheses about the effects of an intervention. The ``Does this work?'' question is best left to the full-scale efficacy trial, and the power calculations for that trial are best based on clinically meaningful differences. Instead, pilot studies should assess the feasibility/acceptability of the approach to be used in the larger study, and answer the ``Can I do this?'' question.  
\end{quote}

This leads to a paradox regarding preliminary studies and power analysis. In order to estimate an effect size, the preliminary study must have a large enough sample size to resolve the question to within commonly accepted statistical error. But in this case, the future study then seems redundant. Example 1 above illustrates this problem.

\subsection{Balance of  false positives and false negatives}\label{sec.balance}

The term  ``publication bias''  can be described as a tendency to favor the reporting and publishing of  statistically significant findings, and is  often characterized as a self-evident flaw in our scientific culture \citep{ easterbrook1991}.  Of course, in practice statistically significant findings tend to be those of greater scientific interest. This is the premise of the publication model represented in Figure \ref{fig.decision.tree}, and the motivation of initiatives such as the OSC-RP.  Viewed this way,   ``publication bias'' is nothing more than type I error, a source of false positives which is inevitable in any experimental science which employs statistical methods. As demonstrated in Section  \ref{sec.prelim}, this leads to an upward bias in effect size, but this is not the result of flawed statistical or experimental methods. 

The complement of ``publicaton bias''  is the  ``file drawer effect'', or the nonpublication of findings with $P > 0.05$, as described  in \cite{rosenthal1979}. We can identify two issues here.  The first is the notion that a study that is unpublished because $P > 0.05$, hence consigned to a file drawer, is a result lost to science, since a null hypothesis $H_o$ is as much scientific fact as is $H_a$. The aggregation of all experimental results concerning a specific scientific question,  whether $H_a$ or $H_o$, published or unpublished, is obviously more informative than experiments considered separately. This is precisely the purpose of the systematic registration of clinical trials (Section \ref{sec.clinical.trials}).

However, one curious feature of much of the discussion of the ``reproducibility crisis'' is its emphasis on the control of false positives at the expense of any recognition of the desirability of also controlling  the false negative rate. As stated in \cite{kraemer2006}:  
\begin{quote}
In the typical small pilot study, the standard error of that effect size is very large. Consequently, there is a substantial probability of underestimation of the effect size, which could lead to inappropriately aborting the study proposal for an RCT. If the study is not aborted, there is a substantial probability of serious overestimation of the effect size, which would lead  to an underpowered study and a failed RCT. 
\end{quote}
In other words, if ``publication bias'' leads to false positives, the ``file drawer effect''  leads to false negatives as part of the same process. If this view is accepted, we simply return to the standard notions of type I and type II error.   In this case, all that remains is the technical problem of calculating these  error rates correctly (Example 2 above). 
 
 
 
\subsection{Is the problem small sample size? Power analysis is a decision problem, not an estimation problem.}
   
Small sample size is frequently implicated in failures of reproducibility \citep{ioannidis2005,arain2010,button2013,westlund2016}.  For example, ``[c]ontrary to tradition, a pilot study does not provide a meaningful effect size estimate for planning subsequent studies due to the imprecision inherent in data from small samples ...'' \citep{leon2011}.

Technically, this is true. The approximate type II errors given by Equation \eqref{eq.beta.2} or \eqref{eq.beta.2.2side} converge stochastically to the nominal value $\beta$ as $n \rightarrow\infty$. However, we cannot really describe this as a small-sample problem, since the corrective of increasing $n$ sufficiently is not available.   The purpose of a power analysis is  to ensure that $\alpha$ and $\beta$ are small, but not negligible, since that would require  a larger sample size than is needed.  Once these quantities are fixed, the only remaining parameter in \eqref{eq.beta.2} or \eqref{eq.beta.2.2side} is  $\eta$, and it is important to note that $n$ plays a role only through this quantity. Thus, a power analysis is more in the nature a decision problem, which is solved by minimizing the sample size subject to type I and type II constraints, the solution to which is to set, in this case,  $\eta = z_\alpha+z_\beta$ or $\eta = z_{\alpha/2}+z_\beta$, depending on the test. 

As shown in Example 2, one solution to the small sample size problem is to recognize the power analysis as a sequential decision problem, which can lead to a satisfactory resolution to the problem. In this way, an accurate power analysis is possible without relying on the accuracy of the effect size estimate. The problem is therefore not one of small sample sizes.

\section{Discussion} 
 
We have developed a simple reproducibility  model (Section \ref{sec.model}) which may be used to formally define the reproducibility rates of  effects of scientific interest reported to be statistically significant. In addition, we have shown how to quantify the bias induced in power analyses by the use of empirically observed effect sizes. These models rely on well-known and widely accepted statistical methodologies.  We have then used these models to revisit the reproducibility rates reported by the OSC-RP \citep{open2015estimating}, and found that these apparently low rates can be predicted by these models. Here, the important point is that these models assume that standard statistical practice is being carried out, and that type I and II error rates are correctly reported. In particular, any discrepancy between reported reproducibility rates and what is presented as the ideal can be explained by two factors:
\begin{enumerate}
\item The authors of \cite{open2015estimating} present $1 - \bar{\beta}^* = 92\%$ as the ideal reproducibility rate, where $\bar{\beta}^*$ is the average nominal type II error among reproduced studies. However, the logical implication of this is that all findings reported as significant are true positives, that is,  $PPV = 1$, which in turn implies that the  population effect prevalence is $\pi = 1$ (Section \ref{sec.prevalence}). Clearly, this cannot be the case, since then all investigated  effects would be known in advance to exist. More to the point, it contradicts the main claim of the OSC-RP, that the number of false positives among scientific effects  reported in the literature as significant is self-evidently low (since if  $\pi = 1$ there can be false negatives, but no false positives).
\item
While the authors of \cite{open2015estimating} acknowledge the possibility that publication bias may lead to overestimated effect sizes, and therefore underestimated power rates (under the protocol used by the OSC-RP), this bias was not accounted for in their reproducibility rate estimates. We have shown that when this is done, significantly lower observed reproducibility rates are predicted in comparison to those obtained when the nominal power estimates are accepted (Section \ref{sec.rprev}).  Furthermore,  this problem has already been described in the literature (Section \ref{sec.prevailing.views}). Thus, it could be argued that the failure of the OSC-RP to anticipate this effect is out of character for an initiative intended to serve as a corrective  to flawed research practices.
\end{enumerate}

Of course, the reproducibility rate reported by the OSC-RP does seem low (35/97 = 36\%, Table \ref{table.rp.endpoint}). However, we were able to show that this value is quite compatible with our model, which relies only on the correct application of standard statistical methods, as well as reasonable assumptions about what the true population effect prevalence $\pi$ really is  (Section \ref{sec.rprev}).  Thus, if deviations from good statistical and experimental practice are not needed to explain the OSC-RP report, then it provides no basis on which to claim the existence of a ``reproducibility crisis'', at least not one that could not have been predicted by traditional statistical methods.  

\subsection{The purpose of experiments is to reduce uncertainty}

Uncertain results are not an inherently negative concept; rather, uncertainty is the only reason that scientific studies are conducted. The goal of a clinical trial is to reduce uncertainty. The idea that reproducibility rates are not perfect, and are in fact significantly less than perfect, is already accepted and anticipated in modern scientific practices. Reproducibility and validation are seen as an essential part of science.

While high reproducibility standards  reduce the rate of false positives, they also increase the rate of false negatives, which \bcx are  conceivably the costlier error. \ec After all, in experimental science, false negatives are \bcx potentially valuable findings that are lost. On the other hand, false positives can be flagged by validation studies. \ec A solution to this problem requires a balance between false positive and false negative rates, and thus the rate of reproducibility can be too high as well as too low.

\subsection{Reproducibility can, and should, be modelled using intuitively relevant parameters}

The main outcome of a reproducibility project such as the OSC-RP is clearly the reproducibility rate, or the proportion of the $N$ studies considered which are successfully reproduced. However,  the dependence of this rate on factors which may vary between studies needs to be acknowledged. It is especially important to note that these factors are not limited to type I and II error rates accepted as a standard. 

An especially dramatic example of this is in the different replication protocols used by the OSC-RP  and the ML-RP. The OSC-RP used one replication per study,  accepting as nominal type I, II errors $\alpha^* = 0.05$ and, on average, $\bar{\beta}^* = 0.08$ for that replication study. In contrast, the ML-RP used 35-36 replications per study, effectively reducing the replication type I, II errors to nearly zero. This may have a significant effect on the reported reproducibility rates, and it is entirely attributable to the replication study protocol (Section  \ref{sec.model.def}). 

Apart from the question of replication study protocol, a baseline reproducibility rate $PPV$ that could be expected in the absence of any systematic flaws in research  practice is easily given by the relationship developed in Section \ref{sec.model.def}: 
$$
\textup{Odds}(PPV)=\bigg(\frac{1-\beta}{\alpha}\bigg)\cdot\textup{Odds}(\pi).
$$
Deviations from the baseline can be related to the various parameters.  In terms of $\alpha$, it is possible that $P$-values are being underestimated, by a lack of multiple testing control, improper cross validation, or other factors. This causes an increase in $\alpha$. In terms of $\beta$, it is possible that the study is underpowered, and thus $(1-\beta)$ decreases. These effects are quite distinct, the former adding more false positives to, the latter keeping more true positives from,  the pool of published findings. The final parameter $\pi$ is, of course, not affected by statistical methodology, but can still vary considerably from, ideally, $\pi = 0.5$  for a well-powered RCT to $\pi = 1/1000$, as estimated in \cite{ioannidis2005} for `` [d]iscovery-oriented exploratory research with massive testing''.

Thus, if it is true that the rate of reproducibility has decreased in recent years, then it is as feasible to explain this as a trend towards exploratory, data-driven research as it is to suggest a deterioration of methodological standards, and if it is the former, then there is no indication that the research is unsound because of it. \bcx If the reproducibility rate is deemed unacceptably low, an  obvious corrective is to increase $\pi$ with an increased emphasis on hypothesis-driven or model-based research questions. If this is not possible, the remaining corrective is to accept a greater role for validation, which seems to us a remarkably elegant and practical solution. Especially appealing  is the precise control over the balance between false positives and false negatives (Section \ref{sec.balance}). After all, we can accept PPV = 30\% for a diagnostic test, because we can always  repeat the test to eliminate false positives. But a NPV less than 100\% represents undiagnosed illness, a much costlier error than the false positive.  \ec
 
\bibliographystyle{natbib}
\bibliography{jrep}

\end{document}